\documentclass[runningheads]{llncs}
\usepackage{array}
\usepackage{graphicx}
\usepackage{url}
\usepackage{makecell}
\usepackage{multirow}
\usepackage[caption=false]{subfig}
\usepackage{float}
\floatstyle{plaintop}
\restylefloat{table}
\usepackage{tabularx}
\usepackage{booktabs}
\usepackage{enumitem}
\usepackage{tikz}
\usepackage[font=small,labelfont=bf]{caption}

\usepackage{adjustbox}
\usepackage{caption} 
\begin{document}
\title{A Systematic Mapping Study in AIOps}
%
%
\author{Paolo Notaro\inst{1,2}\orcidID{0000-0003-3567-864X} \and
Jorge Cardoso\inst{2,3} \and
Michael Gerndt\inst{1}}
\authorrunning{P. Notaro et al.}
%
\institute{Chair of Computer Architecture and Parallel Systems, Technical University of Munich, Germany, \email{\{paolo.notaro@|gerndt@in.\}tum.de} \and
Ultra-scale AIOps Lab, Huawei Munich Research Center, Germany, \email{\{name.surname\}@huawei.com} \and
Department of Informatics Engineering/CISUC, University of Coimbra, Portugal}
\maketitle              
%
\begin{abstract}
IT systems of today are becoming larger and more complex, rendering their human supervision more difficult. Artificial Intelligence for IT Operations (AIOps) has been proposed to tackle modern IT administration challenges thanks to AI and Big Data. However, past AIOps contributions are scattered, unorganized and missing a common terminology convention, which renders their discovery and comparison impractical. In this work, we conduct an in-depth mapping study to collect and organize the numerous scattered contributions to AIOps in a unique reference index. We create an AIOps taxonomy to build a foundation for future contributions and allow an efficient comparison of AIOps papers treating similar problems. We investigate temporal trends and classify AIOps contributions based on the choice of algorithms, data sources and the target components. Our results show a  recent and growing interest towards AIOps, specifically to those contributions treating failure-related tasks (62\%), such as anomaly detection and root cause analysis.
\end{abstract}
%
%
%
\section{Introduction}
Modern society is increasingly dependent on large-scale IT infrastructures. At the same time, the latest IT challenges impose higher levels of reliability and efficiency on computer systems. Because of the large increase in size and complexity of these systems, IT operators are increasingly challenged while performing tedious administration tasks manually. This has sparked in recent years much interest towards the study of self-managing and autonomic computing systems to improve efficiency and responsiveness of IT services. While many static algorithmic solutions have been proposed, these automated solutions often show limitations in terms of adaptiveness and scalability. The presence of large data volumes in different modalities motivates the investigation of intelligent learning systems, able to adapt their behavior to new observations and situations.

Artificial Intelligence for IT Operations (AIOps) investigates the use of Artificial Intelligence (AI) for the management and improvement of IT services. AIOps relies Machine Learning, Big Data, and analytic technologies to monitor computer infrastructures and provide proactive insights and recommendations to reduce failures, improve mean-time-to-recovery (MTTR) and allocate computing resources efficiently \cite{andrew_lerner_aiops_2017}. AIOps offers a wide, diverse set of tools for several applications, from efficient resource management and scheduling to complex failure management tasks such as failure prediction, anomaly detection and remediation \cite{dang_aiops_2019,li_predicting_2020}. However, being a recent and cross-disciplinary field, AIOps is still a largely unstructured research area. The existing contributions are scattered across different conferences and apply different terminology conventions. Moreover, the high number of application areas renders the search and collection of relevant papers difficult. Some previous systematic works only treat single tasks or subareas inside AIOps \cite{kobbacy_ai_2007,mukwevho_toward_2018}. This motivates the need for a complete and updated study of AIOps contributions. 

In this paper, we present in-depth analysis of AIOps to cover for these limitations. We have identified and extracted over 1000 AIOps contributions through a systematic mapping study, enabling us to delineate common trends, problems and tools. First, we provide an in-depth description of the methodology followed in our mapping study (Section \ref{sec:SMSS}), reporting and motivating our planning choices regarding problem definition, search, selection and mapping. Then, we present and discuss the results drawn from our study, including the identification of most common topics, data sources, and target components (Section \ref{sec:results}). Finally, Section \ref{sec:conc} summarizes the outcomes and conclusions treated in this work.

\section{Methodology}
\label{sec:SMSS}

\subsection{Systematic Mapping Studies}
A systematic mapping study (SMS) is a research methodology widely adopted in many research areas, including software engineering \cite{petersen_guidelines_2015}. The ultimate goal of a SMS is to provide an overview of a specific research area, to obtain a set of related papers and to delineate trends present inside such area. Relevant papers are collected via predefined search and selection techniques and research trends are identified using categorization techniques across different aspects of the identified papers, e.g. topic or contribution type. We choose to perform a SMS because we are interested in gathering contributions and obtaining statistical insights about AIOps, such as the distribution of works in different subareas and the presence of temporal trends for particular topics. SMSs have also been shown to increase the effectiveness of follow-up systematic literature reviews \cite{petersen_guidelines_2015}. To this end, we have also used our systematic mapping study to collect references for a survey on failure management in AIOps separately published.

\subsection{Planning}
According to the step outline followed in \cite{petersen_guidelines_2015}, a systematic mapping study is composed of:
\begin{itemize}[noitemsep,topsep=0pt]
    \item \textbf{Formulation}, i.e. express the goals intended for the study through research questions. Equally important is to clearly define the scope of investigation;
    \item \textbf{Search}, i.e. define strategies to obtain a sufficiently high number of papers within the scope of investigation. This comprises the selection of one or more search strategies (database search, manual search, reference search, etc.);
    \item \textbf{Selection} (or screening), i.e. define and apply a set of inclusion/exclusion criteria for identifying relevant papers inside the search result set;
    \item \textbf{Data Extraction and Mapping}, i.e. gather the information required to map the selected papers into predefined categorization scheme(s). Finally, results are presented in graphical form, such as histograms or bubble plots.
\end{itemize}
The next sections illustrate and motivate our choices regarding these four steps for our systematic mapping study in AIOps.

\subsection{Formulation}
The main goal of this mapping study is to identify the extent of past research in AIOps. In particular, we would like to identify a representative set of AIOps contributions which can be grouped based on the similarity of goals, employed data sources and target system components. We also wish to understand the relative distribution of publications within these categories and the temporal implications involved. Formally, we articulate the following research questions:

\begin{description}[topsep=2pt]
\setlength\itemsep{-1em}
\item \textbf{RQ1. What categories can be observed while classifying AIOps contributions in scientific literature?}\\
\item \textbf{RQ2. What is the distribution of papers in such categories?}\\
\item \textbf{RQ3. Which temporal trends can be observed for the field of AIOps?}
\end{description}

In terms of scope, we express the boundaries of AIOps as the union of goals and problems in IT Operations when dealt with AI techniques. To circumvent ambiguity about the term AI, we adopt an inclusive convention where we consider AI both date-driven approaches, such as Machine Learning and data mining, as well as goal-based approaches, such as reasoning, search and optimization approaches. However, we mostly concentrate our efforts on the first category due to its stronger presence and connection to AIOps methodologies (e.g.\ data collection). 
	
\subsection{Search and Selection}
\subsubsection{Selection Criteria}
\label{ssec:selection}
We start illustrating the selection principles beforehand, so that the discussion will appear clearer when we describe our result collection strategy, composed of search and selection altogether. In terms of inclusion criteria, we define only one relevance criterion, based on the main topic of the document. Following from our discussion on scoping such inclusion criterion comprises two necessary conditions:
\begin{itemize}[noitemsep,topsep=0pt]
    \item The document references one or more AI methods. These mentions can either be part of the implementation or as part of its discussion/analysis (e.g. in a survey). Any mention to AI algorithms employed by others (i.e.\ mentioned in the related work section or as baseline comparison) that is not strictly the focus of the document, is not considered valid;
    \item The document applies its concepts to some kind of IT system management. We therefore exclude papers with no specific target domain or with a target domain outside of IT Operations. 
\end{itemize}
In terms of exclusion criteria, we define the following as exclusion rules:
\begin{itemize}[noitemsep,topsep=0pt]
    \item The language of the document is not English;
    \item The document is not accessible online;
    \item The document does not belong to the following categories: scientific article (conference paper, journal article), book, white paper;
    \item The main topic of the document is one of the following: cybersecurity, industrial process control, cyber-physical systems, and optical sensor networks.
\end{itemize}
For the special case of survey and review papers, we consider them relevant as long while carrying out our mapping study, but we then exclude them from our final result set, as these articles are useful to find other connected works through references, but they do not constitute novel contributions to the field.

\subsubsection{Database Search}

\begin{table}[t]
    \resizebox{\textwidth}{!}{\begin{tabular}{|c|c|}
    \hline
    \textbf{AI Keywords} & \textbf{IT Operations Keywords}\\
    \hline
    \makecell{
(``AI'' OR ``artificial intelligence'')\\
``classification''\\
``clustering''\\
``logistic regression''\\
``regression''\\
(``DL'' OR ``deep learning'')\\
(``ML'' OR ``Machine Learning'')\\
(``inference'' OR ``logic'' OR ``reasoning)\\
(``supervised'' OR ``unsupervised'' OR \\``semi-supervised'' OR ``reinforcement'') AND (``learning'')\\
(``support vector machine'' OR ``SVM'')\\
(``tree'' OR ``tree-based'' OR ``trees'' OR ``forest'')\\
((``bayesian'' OR ``neural'') AND ``network'')\\
(((``hidden'' AND ``markov'') OR (``gaussian'' \\AND ``mixture'')) AND ``model'')\\	
((``datacenter'' OR ``data center'') AND ``management'')
}
& \makecell{
(``DevOps'' OR ``site reliability engineering'' \\OR ``SRE'') (``IT operations'')\\
(``anomaly detection'' OR ``outlier detection'')\\
(``cloud computing'')\\
(``cloud'')\\
(``fault detection'' OR ``failure detection'')\\
(``fault localization'' OR ``failure localization'')\\
(``fault prediction'' OR ``failure prediction'')\\
(``fault prevention'' OR ``failure prevention'')\\
(``log'' OR ``logs'' OR ``log analysis'')\\
(``metrics'' OR ``KPI'' OR ``key performance indicator'')\\
(``remediation'' OR ``recovery'')\\
(``root-cause analysis'' OR ``root cause analysis'')\\
(``service desk automation'')\\
(``tracing'' OR ``trace'' OR ``traces'')
}\\
    \hline
\end{tabular}}
\caption{The two keyword sets obtained via PICO used for database search.}
\label{tab:keywords}
\end{table}

For the search process, database search represents the first and most important step, as it aims to provide the highest number of results and perform an initial screening of irrelevant papers. We perform database search in three steps: keywording, query construction and result polling.
For keywording we use the PICO technique presented in \cite{petersen_guidelines_2015} to derive a set of keywords for AI and a set of keywords for IT Operations. The keywords are listed in Table \ref{tab:keywords}.
Then, following our scoping considerations, we construct queries so that they return results where both AI and IT Operations are present. In particular, we apply logic conjunction of keywords across all combinations of the two keyword sets (e.g. ``logistic regression'' and ``cloud computing''). This helps enforcing precision in our search results. For keywords with synonyms and abbreviations, we allow all equivalent expressions via OR disjunction. We also perform general search queries, related to the topic as a whole (e.g. ``AIOps''). Finally, we group some queries with common terms to reduce the number of queries.

We select three online search databases that are appropriate for the scope of investigation: IEEE Xplore, ACM Digital Library and arXiv. For each query we restrict our analysis to the top 2000 results returned. We aggregate results from all searches in one large set of papers, removing duplicates and annotating for each item corresponding search metadata (e.g. number of hits, index position in corresponding searches, etc.).
The result from this step consists of 83817 unique articles. For each item we collect the title, authors, year, publication venue, contribution type and citation count (from Google Scholar).

\subsubsection{Preliminary Filtering and Ranking-based Selection}
In the filtering step we start improving the quality of our selection of papers. First, papers are automatically excluded based on publication venue, for those venues that are clearly irrelevant for topic reasons (e.g. meteorology). We also exclude based on the year of publication (year $< 1990$) as it precedes the advent of large-scale IT services. By doing so, we can exclude approximately 8000 elements.

Usually at this point, a full-text analysis would be performed on all the available papers to screen relevant contributions using the above cited selection rules. Although we partly filtered results, it is still not feasible to perform an exhaustive selection analysis, even as simple as filtering by title. It is also impractical to attempt an automated selection by content, as it is not clear how to perform an efficient, high-recall, high-precision text classification without supervision. Therefore, before proceeding with the rest of the search and selection steps, we apply a ranking procedure on these intermediate results, so that we can prioritize investigation of more relevant papers. We apply the exclusion and inclusion rules of Section \ref{ssec:selection} to the papers examined in ranking order. 

\begin{figure}[t!]%
  \centering
  \captionsetup[subfigure]{position=top, labelfont=bf,textfont=normalfont,singlelinecheck=off,justification=raggedright}
    \subfloat[][\centering]{\includegraphics[width=.47\linewidth]{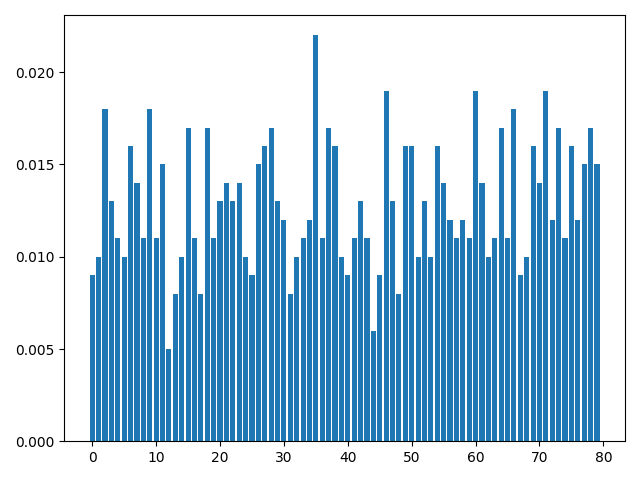}}%
    \qquad
    \subfloat[][\centering]{\includegraphics[width=.47\linewidth]{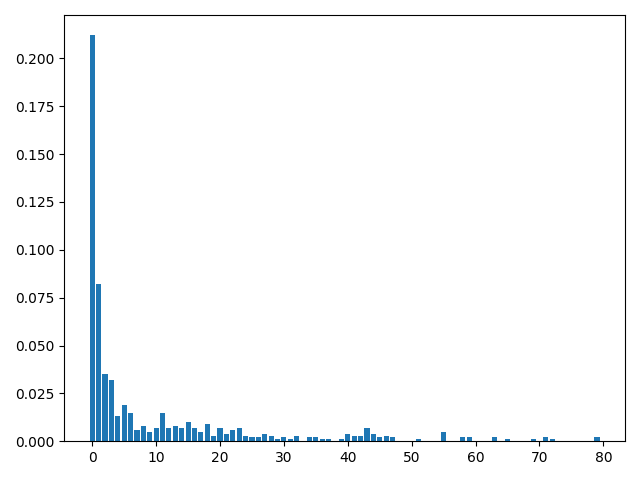}}%
  \caption{Estimated relevance probability for collected papers (y-axis), as a function of the index in the result set (x-axis, in thousands), with paper arranged: (a) in random order (b) using a relevance heuristic based on search hits. We can observe how, thanks to the heuristic (b), the majority of relevant papers can be identified by examining only a small fraction of the set (the top results on the left side).}%
\label{fig:ranking}%
\end{figure}

This approximate procedure however raises the question of when it is convenient to stop our selection and discard the remaining items. To solve this, we develop a new approach from our observations of ranked items. We base the method on the following assumption: a considerable ratio of relevant papers can be identified by ranking and selecting top results using different relevance criteria (conference, position index in the query result set, number of hits in all queries, etc.), but in this sorting scenario we also observe a long-tail distribution of relevant documents, i.e. some relevant papers appear in the last positions even after sorting with our relevance heuristics (see Figure \ref{fig:ranking}). This is coherent with the known impossibility of performing exhaustive systematic literature reviews and mapping studies, as completing the long tail provides less results at the expense of a larger research effort. We assume the ratio of relevant papers in the long tail to be constant and comparable in magnitude to the number of relevant papers when sampled randomly from the result set. Based on this assumption, we proceed as follows:
\begin{itemize}[noitemsep,topsep=0pt]
\item We start screening all papers in the result set, ranked according to different relevance heuristics (e.g. number of hits in queries), and we observe the ratio of relevant papers identified over time;
\item We examine the same papers in random order, and measure the same ratio;
\item When the two ratios are comparable, we assert we reached the tail of the distribution of relevant papers and stop examining and selecting new papers.
\end{itemize}
As sorting criteria, we use the number of hits in the search performed in the previous step, as well as other more complex heuristics, taking into account the index position in result sets and the number of citations. When examining a paper, we look into the full content to identify concepts related to our selection criteria previously illustrated. As done previously with search results, we gather relevant papers in one unique group. Using this stopping criterion, we conclude this selection step when we have identified 430 relevant papers. 

\subsection{Additional Search Techniques}
The ``early stopping'' criterion previously described, while allowing a feasible and comprehensive selection strategy across thousands of contributions, has a natural tendency towards discarding relevant papers. We also expect to miss other relevant papers, not present in the initial set of 83817, because they were not identified by our database search. To cover for these limitations, we apply other search techniques in addition to database search. Differing from before, we here apply our selection criteria exhaustively for each document retrieved.

\subsubsection{Reference Search}
For each of the 430 relevant papers identified in the previous step, we search inside their cited references. In particular, we adopt backward snowball sampling \cite{jalali_systematic_2012}: we include in our relevant set all papers previously cited by a relevant paper whenever they fulfill the selection requirements mentioned above. By doing so, we obtain 631 relevant elements, for a total of 1061.

\subsubsection{Conference Search}
Reference search allows to identify prominent contributions frequently mentioned by other authors. A drawback is the introduction of bias towards specific research groups and authors. We also observed how reference search rewards specific tasks and research fields as they are typically more cited. We therefore apply other search techniques to compensate for these facts. We perform a manual search by inspecting papers published in relevant conferences. These relevant conferences are identified via correlation with other relevant papers and have also been confirmed by experts in the field. We look at the latest 3 editions of each conference, in an effort to compensate the sampling of dated papers performed by reference search. We obtain 5 more papers with this method.

\subsubsection{Iterative Search Improvement}

\begin{table}[t]
    \centering
    \resizebox{\textwidth}{!}{\begin{tabular}{|c|c|c|}
        \hline
        \textbf{k=1}	& \textbf{k=2} & \textbf{k=3}\\
        \hline
        \makecell{
        tpc-w, 1.00 (13)\\
        log-based, 0.92 (11)\\
        sla, 0.84 (48)\\
        stragglers, 0.83 (10)\\
        vm, 0.83 (59)} &
        \makecell{
        defect prediction, 1.00 (34)\\
        (work)load prediction, 1.00 (32)\\
        software aging, 1.00 (13)\\
        resource allocation, 1.00 (6)\\
        hardware failures, 0.89 (8)\\
        } & 
        \makecell{
        software defect prediction, 1.00 (22)\\
        disk failure prediction, 1.00 (8)\\
        failure prediction model, 1.00 (7)\\
        cloud resource provisioning, 1.00 (5)\\
        automatic anomaly detection, 0.88 (7)\\
        }\\
        \hline
    \end{tabular}}%
    \caption{Sample $k$-shingles with relevance probability (and total occurrences).}
    \label{tab:shingles}
\end{table}

To conclude our search, we attempt at improving our initial guess on IT Operations keywords via analysis of the available text content (text and abstract). Using our relevant paper set as positive samples, we perform a statistical analysis to identify $k$-shingles (sets of $k$ consecutive tokens) that appear often in relevant documents (Table \ref{tab:shingles}). In particular, we measure the document relevance probability given the set of shingles observed in the available text content. We choose $k=1, \dots, 5$. We use these shingles as keywords to construct new queries along with previously used AI keywords. We here limit the collection to 20 results per query. Thanks to this step, we identify 20 new relevant papers. As a by-product, we get in contact with frequently cited concepts and keywords in AIOps, later useful for taxonomy and classification.

\vspace{-0.3cm}
\subsection{Data Extraction and Categorization}

After obtaining the result set of relevant papers (counting 1086 contributions), we analyze the available information to draw quantitative results and answer our research questions. We describe here the data extraction process and the analysis techniques employed to gather insights and trends for the AIOps field.

First, we classify the relevant papers according to target components and data sources. Target components indicate a high-level piece of software or hardware in an IT system that the document tries to enhance (e.g. hard drive for hard disk failure prediction). We group components in five high-level categories: code, application, hardware, network and datacenter. Data sources provide an indication of the input information of the algorithm (such as logs, metrics, or execution traces). Data sources are categorized in source code, testing resources, system metrics, key performance indicators (KPIs), network traffic, topology, incident reports, logs and traces. the ``AI Method'' axis denotes the actual algorithm employed, with similar methods aggregated in bigger classes to avoid excessive fragmentation (e.g. ‘clustering’ may contain both k-means and agglomerative hierarchical clustering approaches). Table \ref{tab:st_table} presents a selection of papers from the result set with the corresponding target, source and category annotation.

\begin{table}[t!]
\begin{minipage}[t]{0.5\textwidth}                          
\begin{center}
    
  \begin{adjustbox}{right, width=\textwidth-10pt}        

    \begin{tabular}[t]{|c|c|c|c|c|c|c|c|c|c||c|c|c|c|c|c|} \hline

    \multirow{2}{*}{\textbf{Ref.}} & \multicolumn{9}{c||}{\textbf{Data Sources}} & \multicolumn{5}{c|}{\textbf{Targets}} & \multirow{2}{*}{\textbf{Cat.}}\\\cline{2-15}
     & \rotatebox{90}{\textbf{Source Code}} & \rotatebox{90}{\textbf{Testing Resources}} & \rotatebox{90}{\textbf{System Metrics}} & \rotatebox{90}{\textbf{KPIs/SLO data}} & \rotatebox{90}{\textbf{Network Traffic}} & \rotatebox{90}{\textbf{Topology}} & \rotatebox{90}{\textbf{Incident Reports}} & \rotatebox{90}{\textbf{Event Logs}} & \rotatebox{90}{\textbf{Execution Traces}} & \rotatebox{90}{\textbf{Source Code}} & \rotatebox{90}{\textbf{Application}} & \rotatebox{90}{\textbf{Hardware}} & \rotatebox{90}{\textbf{Network}} & \rotatebox{90}{\textbf{Datacenter}} &\\ \hline
     
    \cite{menzies_data_2007} & $\bullet$ & & & & & & & & & $\bullet$& & & & & 1.1\\ \hline
    
    \cite{natella_fault_2013} & $\bullet$ & $\bullet$ & & & & & & & & $\bullet$ & & & & & 1.2\\ \hline
    
    \cite{garg_methodology_1998} & & & $\bullet$ & & & & & & & & $\bullet$ & & & $\bullet$ & 1.3\\ \hline
    
    \cite{vaidyanathan_comprehensive_2005} & & & $\bullet$ & $\bullet$ & & & & & & & $\bullet$ & & & $\bullet $ & 1.3\\ \hline
    
    \cite{moody_design_2010} & & & & & & & & & & & $\bullet$ & & & $\bullet $ & 1.4\\ \hline
    
    \cite{zheng_long_2017} & & & $\bullet$ & & & & & & & & & $\bullet$ & & & 2.1\\ \hline
    
    \cite{davis_failuresim_2017} & & & $\bullet$ & & & & & & & & & $\bullet$ & & $\bullet$ & 2.1\\ \hline
    
    \cite{costa_system_2014} & & & $\bullet$ & & & & & $\bullet$ & & & & $\bullet$ & & & 2.1\\ \hline
    
    \cite{zhang_syslog_2017} & & & & & & & & $\bullet$ & & & & $\bullet$ & $\bullet$ & & 2.1\\ \hline
    
    \cite{chalermarrewong_failure_2012} & & $\bullet$ & & & & & & & & & $\bullet$ & & & $\bullet$ & 2.2\\ \hline
    
    \cite{cohen_correlating_2004} & & $\bullet$ & $\bullet$ & & & & & & & & $\bullet$ & & & & 2.2\\ \hline
            
    \cite{islam_predicting_2017} & & $\bullet$ & $\bullet$ & & & & & & & & $\bullet$ & & & $\bullet$ & 2.2\\ \hline
    
    \cite{pitakrat_hora_2018} & & $\bullet$ & & & $\bullet$ & & & & & & $\bullet$  & & & $\bullet$ & 2.2\\ \hline
    
    \cite{liang_failure_2007} & & & & & & & & $\bullet$ & & & $\bullet$ & & & $\bullet$ & 2.2\\ \hline
    
    \cite{salfner_using_2007} & & & & & & & & $\bullet$ & & & $\bullet$ & $\bullet$ & $\bullet$ & $\bullet$ & 2.2\\ \hline
    
    \cite{zhang_automated_2016} & & & & & & & & $\bullet$ & & & $\bullet$ & & & & 2.2\\ \hline
    
    \cite{xu_detecting_2009} & $\bullet$ & & & & & & & $\bullet$ & & & & & & $\bullet$ & 3.1\\ \hline
    
    \cite{xu_unsupervised_2018} & & & & $\bullet$ & & & & & & & $\bullet$ & & & & 3.1\\ \hline
    
    \cite{sharma_fault_2013} & & & $\bullet$ & $\bullet$ & & & & & & & $\bullet$ & & & $\bullet$ & 3.1\\ \hline
  
    \cite{lakhina_diagnosing_2004} & & & & & $\bullet$ & $\bullet$ & & & & & & & $\bullet$ & & 3.1\\ \hline
    
    \cite{lakhina_mining_2005} & & & & & $\bullet$ & $\bullet$ & & & & & $\bullet$ & & $\bullet$ & & 3.1\\ \hline
    
    \end{tabular}%

\end{adjustbox}
\end{center}
\end{minipage}%
\begin{minipage}[t]{0.5\textwidth}                          
\begin{center}
  \begin{adjustbox}{left, width=\textwidth-10pt}        
    \begin{tabular}[t]{|c|c|c|c|c|c|c|c|c|c||c|c|c|c|c|c|} \hline
    
        \multirow{2}{*}{\textbf{Ref.}} & \multicolumn{9}{c||}{\textbf{Data Sources}} & \multicolumn{5}{c|}{\textbf{Targets}} & \multirow{2}{*}{\textbf{Cat.}}\\\cline{2-15}
         & \rotatebox{90}{\textbf{Source Code}} & \rotatebox{90}{\textbf{Testing Resources}} & \rotatebox{90}{\textbf{System Metrics}} & \rotatebox{90}{\textbf{KPIs/SLO data}} & \rotatebox{90}{\textbf{Network Traffic}} & \rotatebox{90}{\textbf{Topology}} & \rotatebox{90}{\textbf{Incident Reports}} & \rotatebox{90}{\textbf{Event Logs}} & \rotatebox{90}{\textbf{Execution Traces}} & \rotatebox{90}{\textbf{Source Code}} & \rotatebox{90}{\textbf{Application}} & \rotatebox{90}{\textbf{Hardware}} & \rotatebox{90}{\textbf{Network}} & \rotatebox{90}{\textbf{Datacenter}} &\\ \hline
        
        \cite{du_deeplog_2017} & & & & & & & & $\bullet$ & & & $\bullet$ & & & & 3.1\\ \hline  
        
        \cite{chow_mystery_2014} & & & & & & & & $\bullet$ & $\bullet$ & & $\bullet$ & & & $\bullet$ & 3.1\\ \hline
        
        \cite{barham_magpie_2003} & & & & & & & & & $\bullet$ & & $\bullet$ & & & $\bullet$ & 3.1\\ \hline
        
        \cite{mike_y_chen_path-based_2004} & & & & & & & & & $\bullet$ & & $\bullet$ & & & & 3.1\\ \hline
         
        \cite{moore_internet_2005} & & & & & $\bullet$ & & & & & & & & $\bullet$ & $\bullet$ & 3.2\\ \hline
        
        \cite{zhu_learning_2015} & $\bullet$ & & & & & & & $\bullet$ & & & & & & $\bullet$ & 3.3\\ \hline
        
        \cite{abreu_spectrum-based_2009} & $\bullet$ & $\bullet$ & & & & & & & & $\bullet$ & $\bullet$ & & & & 4.1\\ \hline
        
        \cite{nguyen_fchain_2013} & & & $\bullet$ & & & $\bullet$ & & & & & & $\bullet$ & & $\bullet$ & 4.1\\ \hline
        
        \cite{bahl_towards_2007} & & & & & $\bullet$ & & & & & & & $\bullet$ & $\bullet$ & & 4.1\\ \hline
        
        \cite{yuan_sherlog_2010} & $\bullet$ & & & & & & & $\bullet$ & & & $\bullet$ & & & & 4.2\\ \hline
        
        \cite{attariyan_x-ray_2012} & $\bullet$ & & $\bullet$ & $\bullet$ & & & & & & & $\bullet$ & & & & 4.2\\ \hline
        
        \cite{kandula_shrink_2005} & & & & & $\bullet$ & $\bullet$ & & & & & & & $\bullet$ & & 4.2\\ \hline

        \cite{chen_pinpoint_2002} & & & & & & & & & $\bullet$ & & $\bullet$ & & & $\bullet$ & 4.2\\ \hline
        
        \cite{podgurski_automated_2003} & & $\bullet$ & & & & $\bullet$ & & & & & $\bullet$ & & & & 4.3\\ \hline
        
        \cite{bodik_fingerprinting_2010} & & & $\bullet$ & $\bullet$ & & & & & & & $\bullet$ & & & $\bullet$ & 4.3\\ \hline
        
        \cite{lin_log_2016} & & & & & & & & $\bullet$ & & & $\bullet$ & & & $\bullet$ & 4.3\\ \hline  
        
        \cite{aguilera_performance_2003} & & & & & & & & & $\bullet$ & & $\bullet$ & & & $\bullet$ & 4.3\\ \hline
        
        \cite{shao_efficient_2008} & & & & & & & $\bullet$ & & & & $\bullet$ & & & $\bullet$ & 5.1\\ \hline
        
        \cite{zhou_resolution_2015} & & & & & & & $\bullet$ & & & & $\bullet$ & & & $\bullet$ & 5.2\\ \hline
        
        \cite{lin_hardware_2018} & & & & & & & $\bullet$ & $\bullet$ & & & $\bullet$ & & & $\bullet$ & 5.2\\ \hline
        
        \cite{samir_controller_2019} & & & $\bullet$ & $\bullet$ & & & & & & & $\bullet$ & & & $\bullet$ & 5.3\\ \hline
         
    \end{tabular}
\end{adjustbox}
\end{center}
\end{minipage}%
\vspace{0.05cm}
\begin{adjustbox}{width=1.08\textwidth,center}
\begin{tabular}[t]{|l|l||l|l||l|l|}
    \hline
    \multicolumn{6}{|c|}{\textbf{(Sub)Category Legend}}  \\ \hline
    1.1 & Software Defect Prediction & 2.2 & System Failure Prediction & 4.2 & Root Cause Diagnosis\\ \hline
    1.2 & Fault Injection & 3.1 & Anomaly Detection & 4.3 & RCA - Others\\ \hline
    1.3 & Software Rejuvenation & 3.2 & Internet Traffic Classification & 5.1 & Incident Triage\\ \hline
    1.4 & Checkpointing & 3.3 & Log Enhancement & 5.2 & Solution Recommendation\\ \hline
    2.1 & Hardware Failure Prediction & 4.1 & Fault Localization & 5.3 & Recovery\\ \hline
    \end{tabular}%
\end{adjustbox}%
\captionsetup{font=small}
\caption{Selection of result papers grouped by data sources, targets and (sub)categories.}
\label{tab:st_table}
\end{table}

\begin{figure}[t]
    \centering
    \resizebox{1\textwidth}{!}{\includegraphics{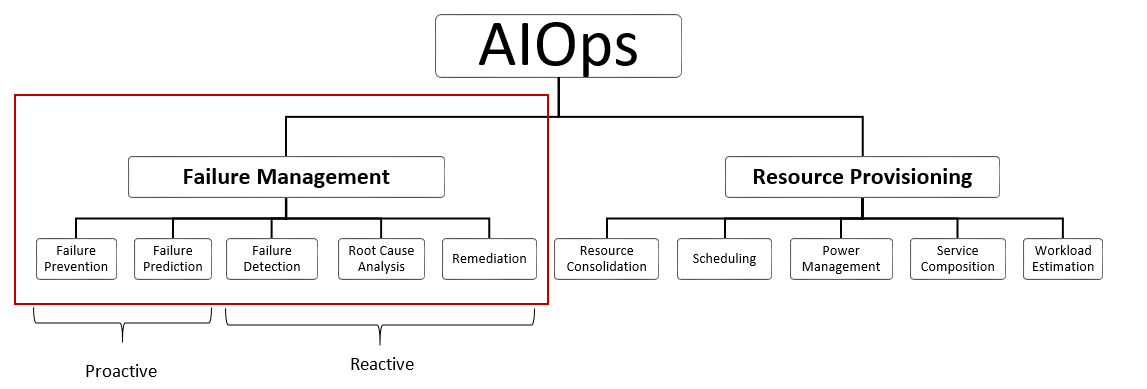}}
    \caption{Taxonomy of AIOps as observed in the identified contributions}
    \label{fig:full-taxonomy}
\end{figure}

Then, we use the result set to infer a taxonomy based on tasks and target goals. The taxonomy is depicted in Figure \ref{fig:full-taxonomy}. We divide in AIOps contributions in failure management (FM), the study on how to deal with undesired behavior in the delivery of IT services; and resource provisioning, the study of allocation of energetic, computational, storage and time resources for the optimal delivery of IT services. Within each of these macro-areas, we further distinguish approaches in categories based on the similarity of goals. In failure management, these categories are failure prevention, online failure prediction, failure detection, root cause analysis (RCA) and remediation. In resource provisioning, we divide contributions in resource consolidation, scheduling, power management, service composition, and workload estimation. We further choose to expand our analysis of FM (red box of Figure \ref{fig:full-taxonomy}) by applying for this macro-area an additional subcategorization based on specific problems. Examples of subcategories are checkpointing for failure prevention, or fault localization for root cause analysis (see also Table \ref{tab:st_table}).

\section{Results}
\label{sec:results}

\begin{figure}[b!]
    \centering
    \resizebox{\textwidth}{!}{\includegraphics{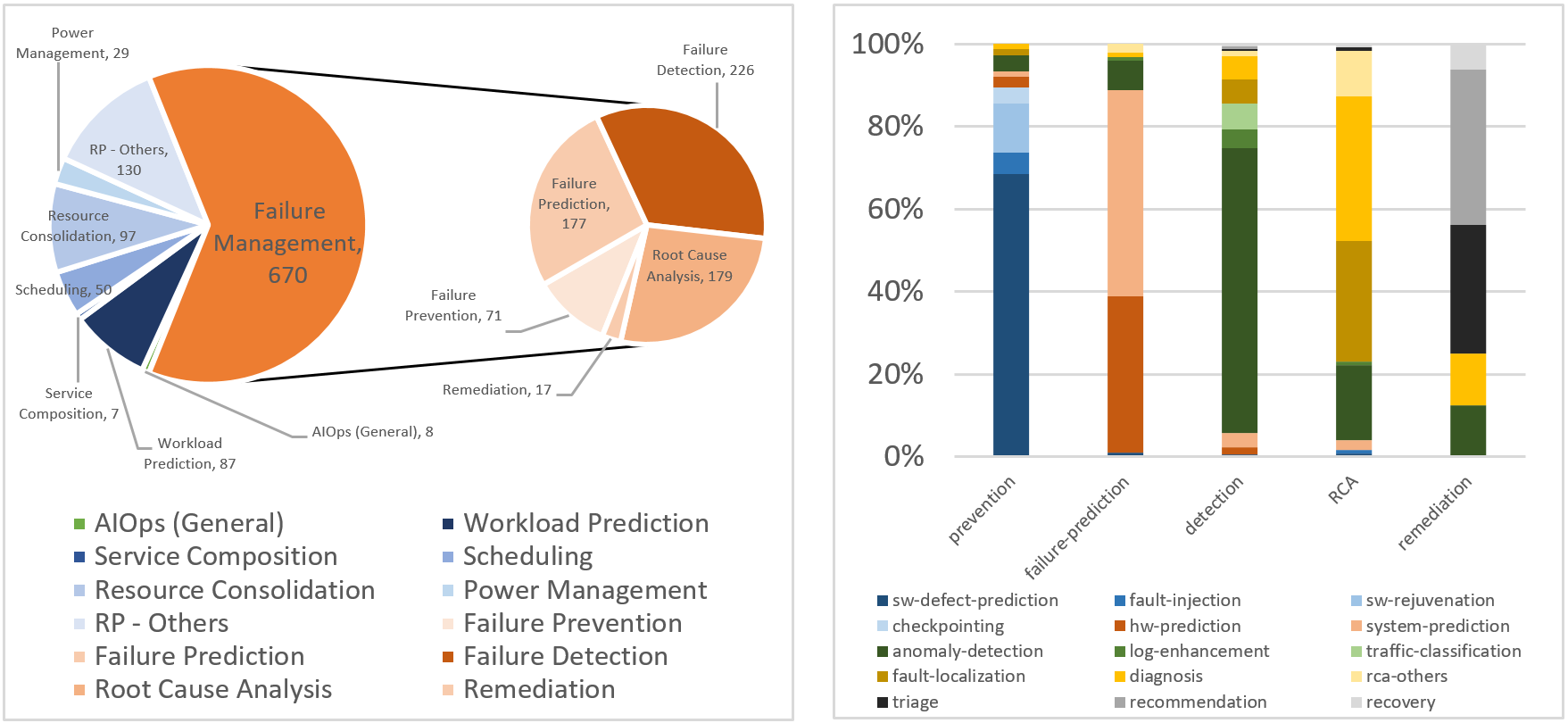}}
    \caption{Left: distribution of AIOps papers in macro-areas and categories. Right: percent distribution of failure management papers by category in corresponding sub-categories.}
    \label{fig:mapping-study_stats}
\end{figure}

 \begin{figure}[t!]
 \hspace{-0.29cm}
    \resizebox{1.043\textwidth}{!}{\includegraphics{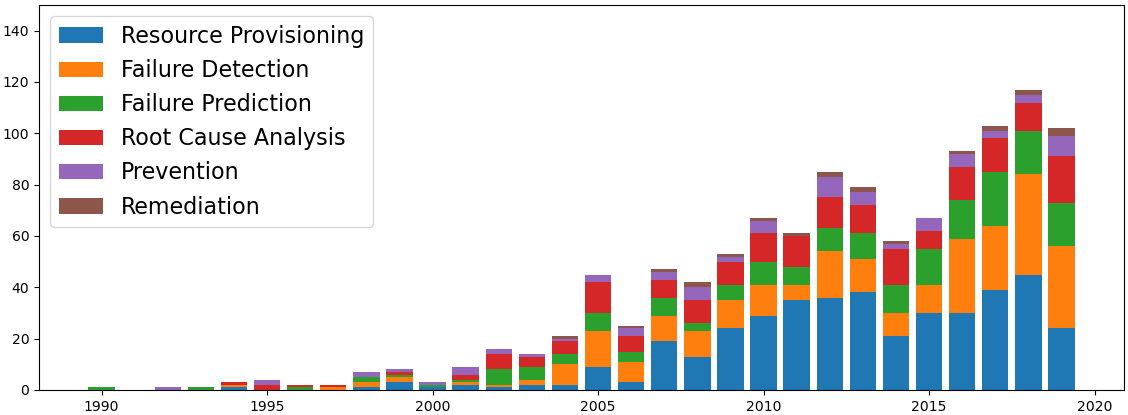}}
    \caption{Published papers in AIOps by year and categories from the described taxonomy.}
    \label{fig:year_category_bar}
\end{figure}

We now discuss the results of our mapping study. We first analyze the distribution of papers in our taxonomy. The left side of Figure \ref{fig:mapping-study_stats} visualizes the distribution of identified papers by macro-area and category. Excluding papers treating AIOps in general (8), we observe that more the majority of items (670, 62.1\%) are associated with failure management (FM), with most contributions concentrated in online failure prediction (26.4\%), failure detection (33.7\%), and root cause analysis (26.7\%); the remaining resource provisioning papers support in large part resource consolidation, scheduling and workload prediction. On the right side, we can observe that the most common problems in FM are software defect prediction, system failure prediction, anomaly detection, fault localization and root cause diagnosis.
To analyze temporal trends present inside the AIOps field, we measured the number of publications in each category by year of publication. The corresponding bar plot is depicted in Figure \ref{fig:year_category_bar}. Overall, we observe a large, on-growing number of publications in AIOps. We can observe how failure detection has gained particular traction in recent years (71 publications for the 2018-2019 period) with a contribution size larger than the entire resource provisioning macro-area (69 publications in the same time frame). Failure detection is followed by root cause analysis (39) and online failure prediction (34), while failure prevention and remediation are the areas with the smallest number of attested contributions (11 and 5, respectively).

\section{Conclusion}

In this paper, we presented our contribution towards better structuring the AIOps field. We planned and conducted a systematic mapping study by means of pre-established formulation, search, selection, and categorization techniques, thanks to which we collected more than 1000 contributions and grouped into several categories thanks to our proposed taxonomy, and differing substantially in terms of goals, data sources and target components. In our result section, we have shown how the majority of papers address failures in different forms. From a time perspective, we observed a generalized on-growing research interest, espcially for tasks such as anomaly detection and root cause analysis.
\label{sec:conc}

%
%
%
\bibliographystyle{splncs04}
\bibliography{bibliography.bib}

\end{document}